\documentclass[longauth]{aa}
\usepackage{graphicx}
\usepackage{txfonts}
\usepackage{xcolor}
\usepackage{hyperref}
\hypersetup{colorlinks, linkcolor=blue, citecolor=blue, urlcolor=blue}
\usepackage{amsmath}
\usepackage{ulem}
\usepackage{soul}
\usepackage{enumitem}

\newcommand{\pievec}{\mbox{\boldmath $\pi$}_{\rm E}}
\newcommand{\muvec}{\mbox{\boldmath $\mu$}}

\newcommand{\te}{t_{\rm E}}
\newcommand{\thetae}{\theta_{\rm E}}

\newcommand{\pien}{\pi_{{\rm E},N}}
\newcommand{\piee}{\pi_{{\rm E},E}}
\newcommand{\dl}{D_{\rm L}}
\newcommand{\ds}{D_{\rm S}}

\definecolor{brown}{rgb}{0.59, 0.29, 0.0}
\definecolor{darkgreen}{rgb}{0.0, 0.42, 0.24}
\definecolor{darkblue}{rgb}{0.01, 0.31, 0.59}
\definecolor{darkblue}{rgb}{0.0, 0.25, 0.42}
\definecolor{blue}{rgb}{0.0,0.0,1.0}
\definecolor{green}{rgb}{0.0,1.0,0.0}

\def\eqalign#1{\null\,\vcenter{\openup\jot
        \ialign{\strut\hfil$\displaystyle{##}$&$
        \displaystyle{{}##}$\hfil \crcr#1\crcr}}\,}

\begin{document}

\title{KMT-2023-BLG-1866Lb: Microlensing super-Earth around an M dwarf host} 
\titlerunning{Microlensing super-Earth KMT-2023-BLG-1866Lb}

\author{
     Cheongho~Han\inst{\ref{inst1}} 
\and Ian~A.~Bond\inst{\ref{inst2}}
\and Andrzej~Udalski\inst{\ref{inst3}} 
\and Chung-Uk~Lee\inst{\ref{inst4}} 
\and Andrew~Gould\inst{\ref{inst5},\ref{inst6}}      
\\
(Leading authors)
\\
     Michael~D.~Albrow\inst{\ref{inst7}}   
\and Sun-Ju~Chung\inst{\ref{inst4}}      
\and Kyu-Ha~Hwang\inst{\ref{inst4}} 
\and Youn~Kil~Jung\inst{\ref{inst4}} 
\and Yoon-Hyun~Ryu\inst{\ref{inst4}} 
\and Yossi~Shvartzvald\inst{\ref{inst8}}   
\and In-Gu~Shin\inst{\ref{inst9}} 
\and Jennifer~C.~Yee\inst{\ref{inst9}}   
\and Hongjing~Yang\inst{\ref{inst10}}     
\and Weicheng~Zang\inst{\ref{inst9},\ref{inst10}}     
\and Sang-Mok~Cha\inst{\ref{inst4},\ref{inst11}} 
\and Doeon~Kim\inst{\ref{inst1}}
\and Dong-Jin~Kim\inst{\ref{inst4}} 
\and Seung-Lee~Kim\inst{\ref{inst4}} 
\and Dong-Joo~Lee\inst{\ref{inst4}} 
\and Yongseok~Lee\inst{\ref{inst4},\ref{inst11}} 
\and Byeong-Gon~Park\inst{\ref{inst4}} 
\and Richard~W.~Pogge\inst{\ref{inst6}}
\\
(The KMTNet Collaboration)
\\
     Fumio~Abe\inst{\ref{inst12}}
\and Ken~Bando\inst{\ref{inst13}}
\and Richard~Barry\inst{\ref{inst14}}
\and David~P.~Bennett\inst{\ref{inst14},\ref{inst15}}
\and Aparna~Bhattacharya\inst{\ref{inst14},\ref{inst15}}
\and Hirosame~Fujii\inst{\ref{inst12}}
\and Akihiko~Fukui\inst{\ref{inst16},}\inst{\ref{inst17}}
\and Ryusei~Hamada\inst{\ref{inst13}}
\and Shunya~Hamada\inst{\ref{inst13}}
\and Naoto~Hamasaki\inst{\ref{inst13}}
\and Yuki~Hirao\inst{\ref{inst18}}
\and Stela~Ishitani Silva\inst{\ref{inst14},\ref{inst19}}
\and Yoshitaka~Itow\inst{\ref{inst12}}
\and Rintaro~Kirikawa\inst{\ref{inst13}}
\and Naoki~Koshimoto\inst{\ref{inst13}}
\and Yutaka~Matsubara\inst{\ref{inst12}}
\and Shota~Miyazaki\inst{\ref{inst20}}
\and Yasushi~Muraki\inst{\ref{inst12}}
\and Tutumi~Nagai\inst{\ref{inst13}}
\and Kansuke~Nunota\inst{\ref{inst13}}
\and Greg~Olmschenk\inst{\ref{inst14}}
\and Cl{\'e}ment~Ranc\inst{\ref{inst21}}
\and Nicholas~J.~Rattenbury\inst{\ref{inst22}}
\and Yuki~Satoh\inst{\ref{inst13}}
\and Takahiro~Sumi\inst{\ref{inst13}}
\and Daisuke~Suzuki\inst{\ref{inst13}}
\and Mio~Tomoyoshi\inst{\ref{inst13}}
\and Paul~J.~Tristram\inst{\ref{inst23}}
\and Aikaterini~Vandorou\inst{\ref{inst14},\ref{inst15}}
\and Hibiki~Yama\inst{\ref{inst13}}
\and Kansuke~Yamashita\inst{\ref{inst13}}
\\
(The MOA Collaboration)
\\
     Przemek~Mr{\'o}z\inst{\ref{inst3}} 
\and Micha{\l}~K.~Szyma{\'n}ski\inst{\ref{inst3}}
\and Jan~Skowron\inst{\ref{inst3}}
\and Rados{\l}aw~Poleski\inst{\ref{inst3}} 
\and Igor~Soszy{\'n}ski\inst{\ref{inst3}}
\and Pawe{\l}~Pietrukowicz\inst{\ref{inst3}}
\and Szymon~Koz{\l}owski\inst{\ref{inst3}} 
\and Krzysztof~A.~Rybicki\inst{\ref{inst3},\ref{inst8}}
\and Patryk~Iwanek\inst{\ref{inst3}}
\and Krzysztof~Ulaczyk\inst{\ref{inst24}}
\and Marcin~Wrona\inst{\ref{inst3}}
\and Mariusz~Gromadzki\inst{\ref{inst3}}          
\and Mateusz~J.~Mr{\'o}z\inst{\ref{inst3}} 
\\
(The OGLE Collaboration)
}

\institute{
      Department of Physics, Chungbuk National University, Cheongju 28644, Republic of Korea\label{inst1}                                                          
\and  Institute of Natural and Mathematical Science, Massey University, Auckland 0745, New Zealand\label{inst2}                                                    
\and  Astronomical Observatory, University of Warsaw, Al.~Ujazdowskie 4, 00-478 Warszawa, Poland\label{inst3}                                                      
\and  Korea Astronomy and Space Science Institute, Daejon 34055, Republic of Korea\label{inst4}                                                                    
\and  Max Planck Institute for Astronomy, K\"onigstuhl 17, D-69117 Heidelberg, Germany\label{inst5}                                                                
\and  Department of Astronomy, The Ohio State University, 140 W. 18th Ave., Columbus, OH 43210, USA\label{inst6}                                                   
\and  University of Canterbury, Department of Physics and Astronomy, Private Bag 4800, Christchurch 8020, New Zealand\label{inst7}                                 
\and  Department of Particle Physics and Astrophysics, Weizmann Institute of Science, Rehovot 76100, Israel\label{inst8}                                           
\and  Center for Astrophysics $|$ Harvard \& Smithsonian 60 Garden St., Cambridge, MA 02138, USA\label{inst9}                                                      
\and  Department of Astronomy and Tsinghua Centre for Astrophysics, Tsinghua University, Beijing 100084, China\label{inst10}                                       
\and  School of Space Research, Kyung Hee University, Yongin, Kyeonggi 17104, Republic of Korea\label{inst11}                                                      
\and  Institute for Space-Earth Environmental Research, Nagoya University, Nagoya 464-8601, Japan\label{inst12}                                                    
\and  Department of Earth and Space Science, Graduate School of Science, Osaka University, Toyonaka, Osaka 560-0043, Japan\label{inst13}                           
\and  Code 667, NASA Goddard Space Flight Center, Greenbelt, MD 20771, USA\label{inst14}                                                                           
\and  Department of Astronomy, University of Maryland, College Park, MD 20742, USA\label{inst15}                                                                   
\and  Department of Earth and Planetary Science, Graduate School of Science, The University of Tokyo, 7-3-1 Hongo, Bunkyo-ku, Tokyo 113-0033, Japan\label{inst16}  
\and  Instituto de Astrof{\'i}sica de Canarias, V{\'i}a L{\'a}ctea s/n, E-38205 La Laguna, Tenerife, Spain\label{inst17}                                           
\and  Institute of Astronomy, Graduate School of Science, The University of Tokyo, 2-21-1 Osawa, Mitaka, Tokyo 181-0015, Japan\label{inst18}                       
\and  Oak Ridge Associated Universities, Oak Ridge, TN 37830, USA\label{inst19}                                                                                    
\and  Institute of Space and Astronautical Science, Japan Aerospace Exploration Agency, 3-1-1 Yoshinodai, Chuo, Sagamihara, Kanagawa 252-5210, Japan\label{inst20} 
\and  Sorbonne Universit\'e, CNRS, UMR 7095, Institut d'Astrophysique de Paris, 98 bis bd Arago, 75014 Paris, France\label{inst21}                                 
\and  Department of Physics, University of Auckland, Private Bag 92019, Auckland, New Zealand\label{inst22}                                                        
\and  University of Canterbury Mt.~John Observatory, P.O. Box 56, Lake Tekapo 8770, New Zealand\label{inst23}                                                      
\and  Department of Physics, University of Warwick, Gibbet Hill Road, Coventry, CV4 7AL, UK\label{inst24}                                                          
}                                                                                                                                                       
\date{Received ; accepted}

\abstract
{}
{
We investigate the nature of the short-term anomaly that appears in the lensing light curve of
KMT-2023-BLG-1866. The anomaly was only partly covered due to its short duration, less than a day, 
coupled with cloudy weather conditions and restricted nighttime duration. 
}
{
Considering intricacy of interpreting partially covered signals, we thoroughly explore all
potential degenerate solutions. Through this process, we identify three planetary scenarios 
that equally well account for the observed anomaly. These scenarios are characterized by the 
specific planetary parameters: 
$(s, q)_{\rm inner} = [0.9740 \pm 0.0083, (2.46 \pm 1.07) \times 10^{-5}]$, 
$(s, q)_{\rm intermediate} = [0.9779 \pm 0.0017, (1.56 \pm 0.25)\times 10^{-5}]$, and 
$(s, q)_{\rm outer} = [0.9894 \pm 0.0107, (2.31 \pm 1.29)\times 10^{-5}]$,
where $s$ and $q$ denote the projected separation (scaled to the Einstein radius) and mass ratio 
between the planet and its host, respectively. We identify that the ambiguity between the inner 
and outer solutions stems from the inner-outer degeneracy, while the similarity between the 
intermediate solution and the others is due to an accidental degeneracy caused by incomplete 
anomaly coverage.  
}
{
Through Bayesian analysis utilizing the constraints derived from measured lensing observables
and blending flux, our estimation indicates that the lens system comprises a very low-mass
planet orbiting an early M-type star situated approximately (6.2 -- 6.5)~kpc from Earth  in terms 
of median posterior values for the different solutions.  The median mass of the planet host is in 
the range of (0.48 -- 0.51)~$M_\odot$, and that of the planet's mass spans a range of (2.6 -- 
4.0)~$M_{\rm E}$, varying across different solutions.  The detection of KMT-2023-BLG-1866Lb signifies 
the extension of the lensing surveys to very low-mass planets that have been difficult to be detected 
from earlier surveys.
}
{}


\maketitle

\begin{table*}[t]
\caption{Additional lensing events with planet-to-host mass ratios $q < 10^{-4}$.\label{table:one}}
\begin{tabular}{lllllll}
\hline\hline
\multicolumn{1}{c}{Parameter}         &
\multicolumn{1}{c}{KMTNet reference}  &
\multicolumn{1}{c}{$\log q$}          &
\multicolumn{1}{c}{Reference}        \\
\hline
 KMT-2023-BLG-1431    &  --                   &   -$4.14 \pm 0.10$           &    \citet{Bell2023}        \\ 
 OGLE-2018-BLG-0677   &  KMT-2018-BLG-0816    &   -$4.11 \pm 0.10$           &    \citet{Herrera2020}     \\ 
 KMT-2021-BLG-1391    &  --                   &   -$4.4 \pm 0.18 $           &    \citet{Ryu2022}         \\ 
 KMT-2021-BLG-0171    &  --                   &   -$4.3$ or $-4.7$           &    \citet{Yang2022}        \\ 
 MOA-2022-BLG-249     &  KMT-2022-BLG-0874    &   $\sim$ -4.1                &    \citet{Han2023a}        \\ 
 KMT-2022-BLG-0440    &  --                   &   -$4.4 \pm 0.18$            &    \citet{Zhang2023}       \\ 
 KMT-2020-BLG-0414    &  --                   &   $\sim$ -4.95               &    \citet{Zang2021b}       \\ 
 KMT-2021-BLG-0912    &  --                   &   $\sim$ -4.56 or -4.95      &    \citet{Han2022b}        \\ 
\hline                                                                                                                 
\end{tabular}
\end{table*}

\section{Introduction}\label{sec:one}
                    
Due to its independence from the luminosity of lensing objects, microlensing was initially 
suggested as a means to detect dark matter in the form of compact objects lying in the Galactic 
halo \citep{Paczynski986}. This concept spurred the initiation of first-generation lensing 
surveys in the 1990s: Massive Astrophysical Compact Halo Object \citep[MACHO:][]{Alcock1993}, 
Optical Gravitational Lensing Experiment \citep[OGLE:][]{Udalski1994}, and Exp\'erience pour 
la Recherche d'Objets Sombres \citep[EROS:][]{Aubourg1993}.  Studies on the lensing behavior 
of events involving binary lens objects have expanded the scope of lensing to encompass planet 
detection \citep{Mao1991, Gould1992b}. The first microlensing planet was reported in 2003 
\citep{Bond2004}, nearly a decade after the commencement of lensing experiments. A planet's 
signal within a lensing light curve manifests as a brief deviation, lasting several hours for 
Earth-mass planets and several days even for giant planets. The delay in planet detection 
stemmed from the fact that the early-generation experiments primarily geared towards dark 
matter detection, and thus the observational cadence of these surveys, roughly a day, fell 
short for effective planet detections. To meet the required observational cadence for planet 
detections, planetary lensing experiments conducted during the period from the mid-1990s to 
the mid-2010s adopted a hybrid strategy. In this setup, survey groups focused on identifying 
lensing events, while follow-up teams densely observed a limited subset of detected events 
using multiple narrow-field telescopes.

Starting in the mid-2010s, planetary microlensing experiments transitioned to a new stage. This
involved a significant boost in the frequency of observations within lensing surveys, achieved by
utilizing multiple telescopes across the globe, all equipped with extensive wide-field cameras.
Currently, three groups are carrying out lensing surveys including the Microlensing Observations 
in Astrophysics survey \citep[MOA:][]{Bond2001}, the Optical Gravitational Lensing Experiment~IV 
\citep[OGLE-IV:][]{Udalski2015}, and the Korea Microlensing Telescope Network 
\citep[KMTNet:][]{Kim2016}. The observational cadence of these surveys reaches down to 0.25~hr, 
which is nearly two orders higher than the cadence of early-generation experiments. The significant 
boost in observational cadence has resulted in a remarkable rise in the detection rate of events, 
with current observations detecting over 3000 events compared to dozens in the early experiments.  
The enhanced ability to closely monitor all lensing events has substantially increased the detection 
rate of planets as well, with present surveys averaging approximately 30 planet detections per year 
\citep{Gould2022}. As a result, microlensing has now emerged as the third most productive method for 
planet detection, following behind the transit and radial velocity techniques.

The most significant rise in detection rates among identified microlensing planets is notably
observed for those having planet-to-host mass ratios below $q < 10^{-4}$. As the mass ratio diminishes,
the duration of a planet's signal becomes shorter. Consequently, only six planets with $q < 10^{-4}$ 
had been identified during the initial 13 years of lensing surveys. However, with the full
implementation of high-cadence surveys, the detection rate for these planets experienced a
substantial surge, resulting in the identification of 32 planets within the 2016--2023 time frame.
Table~14 in \citet{Zang2023} summarizes the information on the 24 planets\footnote{
KMT-2016-BLG-0212 \citep{Hwang2018a} , KMT-2016-BLG-1105 \citep{Zang2023}, OGLE-2016-BLG-1195
\citep{Shvartzvald2017}, KMT-2017-BLG-1003 \citep{Zang2023}, KMT-2017-BLG-0428 \citep{Zang2023},
KMT-2017-BLG-1194 \citep{Zang2023}, OGLE-2017-BLG-0173 \citep{Hwang2018b}, OGLE-2017-BLG-1434
\citep{Udalski2018}, OGLE-2017-BLG-1691 \citep{Han2022c}, OGLE-2017-BLG-1806 \citep{Zang2023},
KMT-2018-BLG-0029 \citep{Gould2020}, KMT-2018-BLG-1025 \citep{Han2021}, KMT-2018-BLG-1988
\citep{Han2022a}, OGLE-2018-BLG-0506 \citep{Hwang2022}, OGLE-2018-BLG-0532 \citep{Ryu2020},
OGLE-2018-BLG-0977 \citep{Hwang2022}, OGLE-2018-BLG-1185 \citep{Kondo2021}, 
OGLE-2018-BLG-1126 \citep{Gould2022}, KMT-2019-BLG-0253 \citep{Hwang2022}, KMT-2019-BLG-0842
\citep{Jung2020}, KMT-2019-BLG-1367 \citep{Zang2023}, KMT-2019-BLG-1806 \citep{Zang2023},
OGLE-2019-BLG-0960 \citep{Yee2021}, and OGLE-2019-BLG-1053 \citep{Zang2021a}.} discovered 
between 2016 and 2019, while Table~\ref{table:one} contains information on the 8 planets 
identified thereafter.

In this study, we present the discovery of a very low-mass-ratio planet uncovered from the
microlensing surveys conducted in the 2023 season. The presence of the planet was revealed
through a partially covered dip anomaly feature appearing near the peak of a lensing light 
curve.  Through comprehensive analysis, we ascertain that this signal indeed originates from 
a planetary companion characterized by a mass ratio $q < 10^{-4}$.

\section{Observation and data}\label{sec:two}

The planet was identified from the analysis of the lensing event KMT-2023-BLG-1866. The source
of the event, with a baseline magnitude of $I_{\rm base}=18.19$, resides toward the Galactic bulge 
field at the equatorial coordinates $({\rm RA}, {\rm Dec})_{\rm J2000} = $ (18:13:55.74, $-$28:26:50.60), 
corresponding to the Galactic coordinates $(l, b) = (3^\circ \hskip-2pt .5090, -5^\circ \hskip-2pt 
.1468)$.  The brightening of the source via lensing was first recognized by the KMTNet group on July 
31, 2023, which corresponds to the abridged heliocentric Julian date ${\rm HJD}^\prime \equiv 
{\rm HJD}-2460000 \sim 156$.  The KMTNet group conducts its lensing survey using three identical 
telescopes, each featuring a 1.6-meter aperture and a camera capable of capturing a 4 square-degree 
field. To ensure continuous monitoring of lensing events, these telescopes are strategically 
positioned across three Southern Hemisphere countries. The locations of the individual telescopes 
are the Siding Spring Observatory in Australia (KMTA), the Cerro Tololo Inter-American Observatory 
in Chile (KMTC), and the South African Astronomical Observatory in South Africa (KMTS).

The source of the event is also situated within the regions of the sky covered by the other two
lensing surveys conducted by the MOA and OGLE groups. The OGLE group carries out its survey
with the use of a 1.3-meter telescope lying at the Las Campanas Observatory in Chile. The camera
mounted on the OGLE telescope offers a field of view spanning 1.4 square degrees.  The MOA
group performs its survey with a 1.8-meter telescope positioned at the Mt. John Observatory in
New Zealand. The MOA telescope is equipped with a camera capturing a 2.2 square-degree area of
the sky in a single exposure. The OGLE team identified the event on August 13 (HJD$^\prime$ = 169) 
and designated it as OGLE-2023-BLG-1093. Later, on September 25 (HJD$^\prime$ = 212), the MOA team
spotted the same event, labeling it as MOA-2023-BLG-438. Hereafter, we assign the event the
designation KMT-2023-BLG-1866, using the identification reference from the initial detection 
survey group. The KMTNet and OGLE surveys conducted their primary observations of the event in 
the $I$ band, whereas the MOA survey utilized its customized MOA-$R$ band for observations. Across
all surveys, a subset of images was obtained in the $V$ band specifically for measuring the color
of the source star.

The data of the event were processed using the photometry pipelines that are customized to the
individual survey groups: KMTNet employed the \citet{Albrow2009} pipeline, OGLE utilized the
\citet{Udalski2003} pipeline, and MOA employed the \citet{Bond2001} pipeline. For the use of the
optimal data, the KMTNet data set used in our analyses was refined through a re-reduction
process using the code developed by \citet{Yang2024}. For each data set, error bars estimated
from the photometry pipelines were recalibrated not only to ensure consistency of the error bars
with the scatter of data but also to set the $\chi^2$ value per degree of freedom (d.o.f.) for 
each data set to unity. This normalization process was done in accordance with the procedure 
outlined by \citet{Yee2012}.\footnote{The photometry data are avaialable through the web page: 
{http://astroph.chungbuk.ac.kr/$\sim$cheongho/KMT-2023-BLG-1866/data.html}.}.

\begin{figure}[t]
\includegraphics[width=\columnwidth]{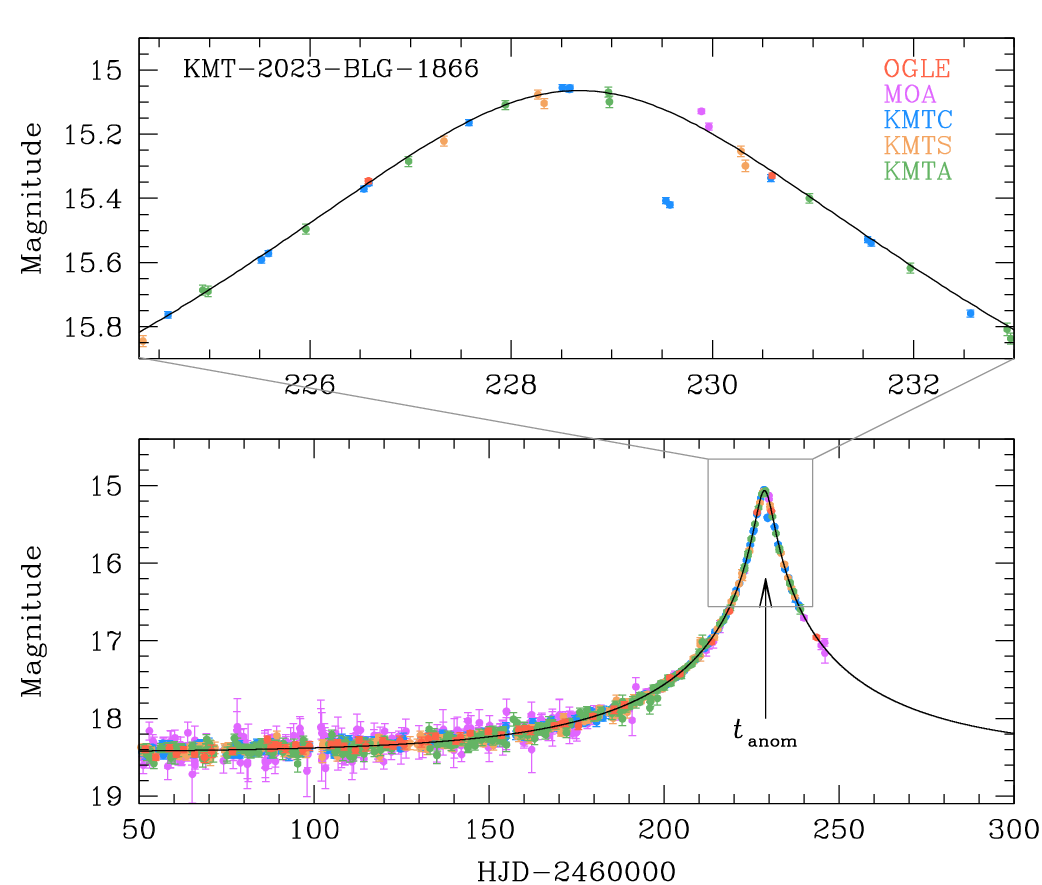}
\caption{
Light curve of KMT-2023-BLG-1866. The lower panel provides a comprehensive view of the
light curve, while the upper panel offers a magnified view of the top region. The color 
of each data point corresponds to the respective dataset indicated in the legend.  The 
curve drawn over data points is a single-lens single source model.
}
\label{fig:one}
\end{figure}

Figure~\ref{fig:one} illustrates the lensing light curve compiled from the collective data of the 
three lensing surveys. The lower panel displays the entire view, while the upper panel provides 
a zoomed-in perspective around the peak. We note that the observational season concluded at 
HJD$^\prime$ = 245, hence data beyond that point are unavailable. The event reached its peak at 
${\rm HJD}^\prime \sim 228.6$ with a moderately high magnification $A_{\rm peak}\sim 32$. Upon 
close examination of the peak region in the light curve, we identified an anomaly centered around 
${\rm HJD}^\prime \sim 229.5$, indicated by an arrow labeled "$t_{\rm anom}$" in the lower panel. 
The anomaly, captured by two KMTC data points, exhibits a negative deviation from the single-lens 
single-source (1L1S) curve that is drawn over the data points. Besides these data points, the two 
MOA points taken at  ${\rm HJD}^\prime=229.89$ and 229.97 exhibits slight positive deviations. The 
anomaly was only partly covered due to its short duration, less than a day, coupled with cloudy 
weather conditions at the MOA site on the day  ${\rm HJD}^\prime=228$ and the KMTS and KMTA sites 
on the day  ${\rm HJD}^\prime=229$, as well as the restricted nighttime duration toward the 
conclusion of the bulge season.

\section{Light curve modeling}\label{sec:three}

Short-term central anomalies in the light curve of a high-magnification event can stem from 
three primary causes. Firstly, they may arise due to the presence of a planetary companion 
orbiting the lens, positioned approximately at the Einstein ring of the primary lens 
\citep{Griest1998}.  Secondly, such anomalies can also result from a binary companion to the 
lens, in which the separation is notably different from the Einstein radius $\thetae$, being 
either significantly smaller or larger \citep{An2002}. A faint binary companion to the source 
can also create a transient central anomaly \citep{Gaudi1998}. However, this particular channel 
results in a positive anomaly, prompting us to exclude this specific scenario from consideration.

\begin{table*}[t]
\caption{Lensing parameters of three local solutions.\label{table:two}}
\begin{tabular}{lllllll}
\hline\hline
\multicolumn{1}{c}{Parameter}         &
\multicolumn{1}{c}{Inner}             &
\multicolumn{1}{c}{Intermediate}      &
\multicolumn{1}{c}{Outer}             \\
\hline
 $\chi^2$/d.o.f.            &   $1759.47/1763        $    &   $1759.05/1763        $    &  $1759.16/1763       $       \\ 
 $t_0$ (HJD$^\prime$)       &   $228.660  \pm 0.012  $    &   $228.649  \pm 0.012  $    &  $228.654  \pm 0.012 $      \\ 
 $u_0$ (10$^{-2}$)          &   $  3.29   \pm 0.16   $    &   $  3.19   \pm 0.16   $    &  $  3.32   \pm 0.16  $      \\ 
 $\te$ (days)               &   $ 75.03   \pm 3.40   $    &   $ 77.30   \pm 3.55   $    &  $ 74.22   \pm 3.48  $      \\ 
 $s$                        &   $  0.9740 \pm 0.0083 $    &   $  0.9779 \pm 0.0017 $    &  $  0.9894 \pm 0.0107$      \\ 
 $q$ (10$^{-5}$)            &   $  2.46   \pm 1.07   $    &   $  1.56   \pm 0.25   $    &  $  2.31   \pm 1.29  $      \\ 
 $\alpha$ (rad)             &   $  1.2187 \pm 0.0065 $    &   $  1.2135 \pm 0.0076 $    &  $  1.2184 \pm 0.0068$      \\ 
 $\rho$ (10$^{-3}$)         &   $  2.28   \pm 0.37   $    &   $  1.99   \pm 0.18   $    &  $  2.26   \pm 0.61  $      \\ 
 $\pien$                    &   $  0.089  \pm 0.102  $    &   $  0.069  \pm 0.097  $    &  $  0.024  \pm 0.097 $      \\
 $\piee$                    &   $ -0.037  \pm 0.0294 $    &   $ -0.064  \pm 0.031  $    &  $ -0.038  \pm 0.030 $      \\
 $ds/dt$ (yr$^{-1}$)        &   $  0.13   \pm 0.66   $    &   $  0.44   \pm 0.63   $    &  $  0.31   \pm 0.76  $      \\
 $d\alpha/dt$ (yr$^{-1}$)   &   $  0.82   \pm 0.71   $    &   $ -1.33   \pm 0.80   $    &  $ -1.21   \pm 0.77  $      \\
                                          
\hline                                                                                                                            
\end{tabular}                             
\tablefoot{ ${\rm HJD}^\prime = {\rm HJD}- 2460000$.  }
\end{table*}

To interpret the anomaly, we employed a binary-lens single-source (2L1S) model to analyze the
light curve. The modeling aimed to find a set of lensing parameters (solution) that most accurately
describe the observed light curve. Under the assumption that the relative motion between the lens
and the source is rectilinear, a 2L1S lensing light curve is described by seven fundamental lensing
parameters. The initial subset of these parameters $(t_0, u_0, \te)$ characterizes the approach of 
the source to the lens. Each parameter denotes the time of closest approach, the separation between
the lens and source at that specific time (impact parameter), and the event timescale. The event
time scale is defined as the time required for the source to traverse the Einstein radius, that is,
$\te=\thetae/\mu$, where $\mu$ denotes the relative lens-source proper motion. The subsequent set 
of parameters $(s, q, \alpha)$ describe the binary-lens system itself and the direction of the 
source's approach to the lens. These parameters represent the projected separation and mass ratio 
between the binary lens components $M_1$ and $M_2$, and the angle between the axis formed by $M_1$ 
and $M_2$ and the direction of $\muvec$ vector, respectively. The lengths of $u_0$ and $s$ are 
scaled to $\thetae$. Finally, the parameter $\rho$ is defined as the ratio of the angular source 
radius $\theta_*$ to the Einstein radius, that is, $\rho = \theta_*/\thetae$ (normalized source 
radius). This parameter characterizes the finite-source magnifications during instances when the 
source crosses or closely approaches lens caustics.

\begin{figure}[t]
\includegraphics[width=\columnwidth]{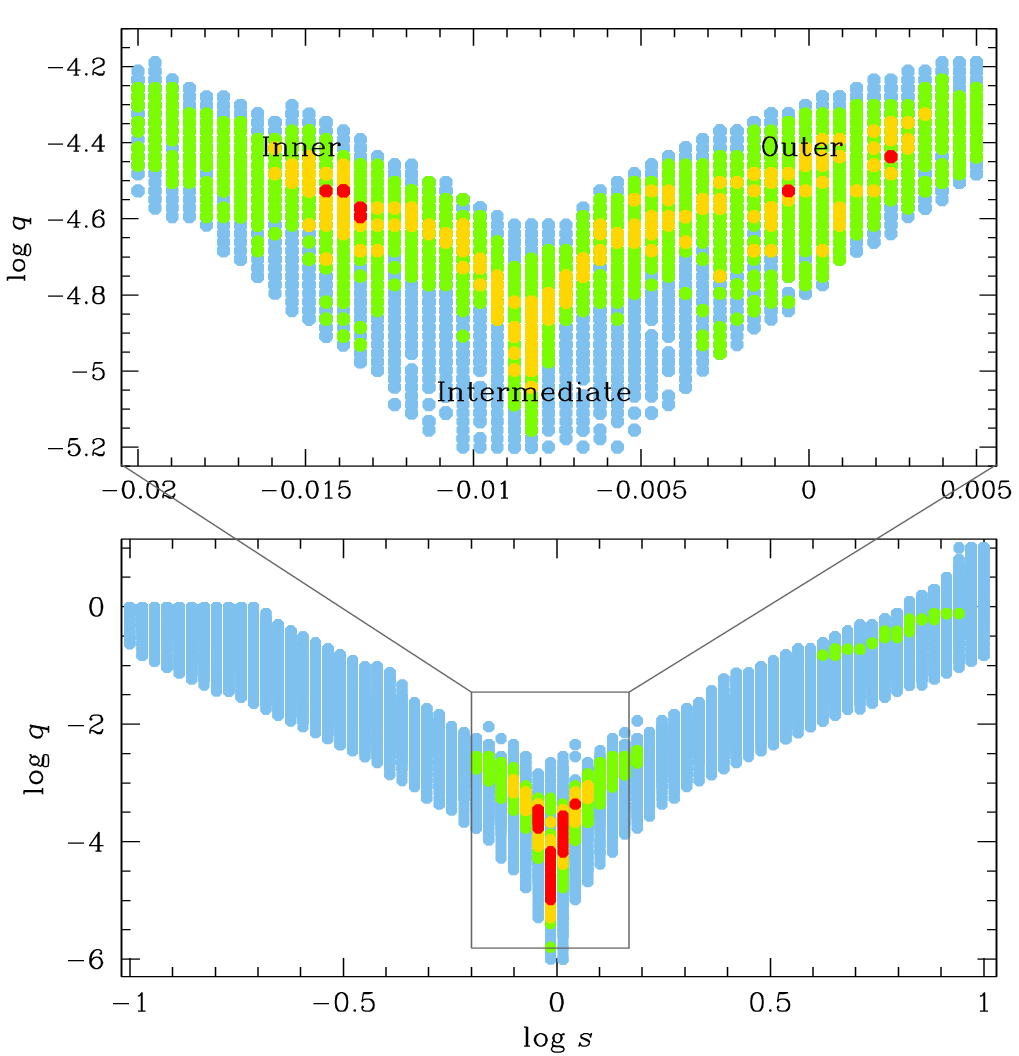}
\caption{
$\Delta\chi^2$ map on the $(\log s, \log q)$ parameter plane obtained from the grid search.
In the lower panel, the entire grid of the inspected region is displayed, while the upper 
panel provides a closer look at the area surrounding the three local solutions.  Three 
identified local solutions are labeled as "inner", "intermediate", and "outer".  The color 
scheme is configured to represent points as red ($< 1n$), yellow ($< 4n$), green ($< 9n$), 
and cyan ($< 16n$). Here $n=1$ for the upper map and $n=10$ for the lower map.
}
\label{fig:two}
\end{figure}

From the analyses of partially covered short-term central anomalies in three lensing events 
KMT-2021-BLG-2010, KMT-2022-BLG-0371, and KMT-2022-BLG-1013, \citet{Han2023b} recently showcased 
the intricacy of interpreting these signals due to the existence of multiple solutions affected 
by various types of degeneracies. To explore all potential degenerate solutions, our modeling 
approach commenced with grid searches for the binary lensing parameters $s$ and $q$ with multiple 
starting values of $\alpha$.   
During this process, we iteratively minimized $\chi^2$ to determine the remaining lensing 
parameters.  We achieved this using the Markov Chain Monte Carlo (MCMC) method, employing an 
adaptive step size Gaussian sampler detailed in \citet{Doran2004}.  Interpreting a central 
anomaly observed in a high-magnification event can be complicated by the potential degeneracy 
between binary and planetary interpretations as demonstrated by \citet{Choi2012}.  To explorer 
such degeneracies, we broaden the range for $s$ and $q$ to encompass both binary and planetary 
scenarios: $-1.0< \log s < 1.0$ and $-6.0< \log q < 1.0$.  The parameter space was partitioned 
into $70 \times 70$ grids, with 21 initial values for the source trajectory angle $\alpha$ evenly 
distributed within the range of $0 < \alpha \leq 2\pi$.  For the computation of finite-source 
magnifications, we utilized the map-making approach detailed by \citet{Dong2006}. In our computations, 
we accounted for limb-darkening effects by modeling the surface brightness variation of the source 
as $S \propto 1 - \Gamma (1-3 \cos \phi/2)$, where $\Gamma$ represents the limb-darkening coefficient, 
and $\phi$ denotes the angle between the line extending from the source center to the observer and 
the line extending from the source center to the surface point.  As will be discussed in 
Sect.~\ref{sec:four}, the source star of the event is identified as a turnoff star of a mid G spectral 
type. We adopted an $I$-band limb-darkening coefficient of $\Gamma_I = 0.45$ from \citet{Claret2000}, 
under the assumptions of an effective temperature of $T_{\rm eff} = 5500$~K, a surface gravity of 
$\log(g/g_\odot) = -1.0$, and a turbulence velocity of $v_{\rm turb} = 2$ km/s.  Following the 
identification of local solutions on the $\Delta\chi^2$ map in the $\log s$--$\log q$ plane, we 
proceeded to refine the lensing parameters by gradually narrowing down the parameter space.

\begin{figure}[t]
\includegraphics[width=\columnwidth]{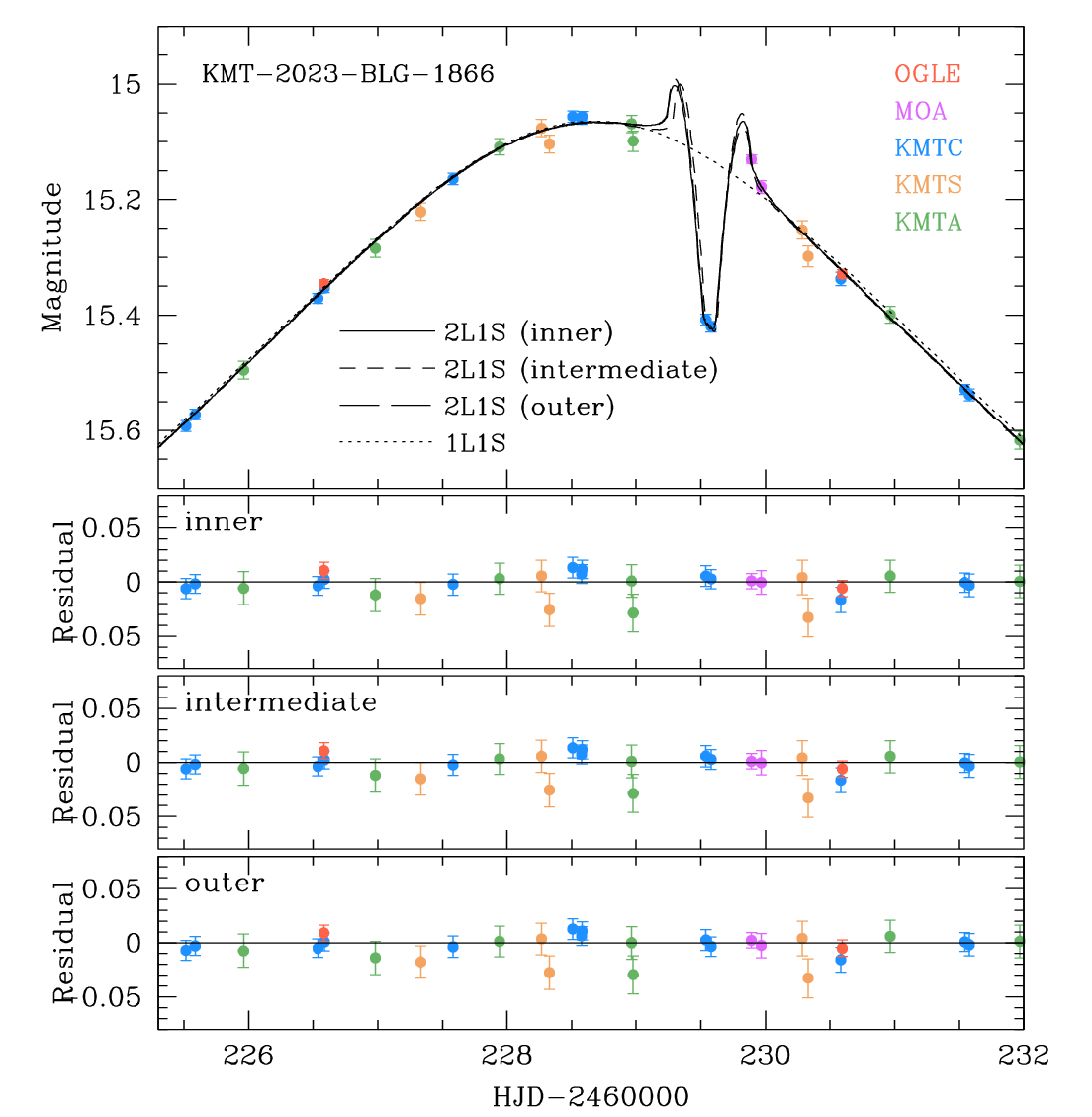}
\caption{
Comparison of the models curves of the three identified local solutions ("inner", "intermediate",
and "outer" solutions) in the region of the anomaly.  The model curves of the the solutions are
drawn over the data points in different line types.  The lower three panels shows the residuals 
from the individual solutions.  The lensing parameters of the individual solutions are listed in 
Table~\ref{table:two}, and the corresponding lens-system configurations are presented in 
Fig.~\ref{fig:four}.
}
\label{fig:three}
\end{figure}

\begin{figure}[t]
\includegraphics[width=\columnwidth]{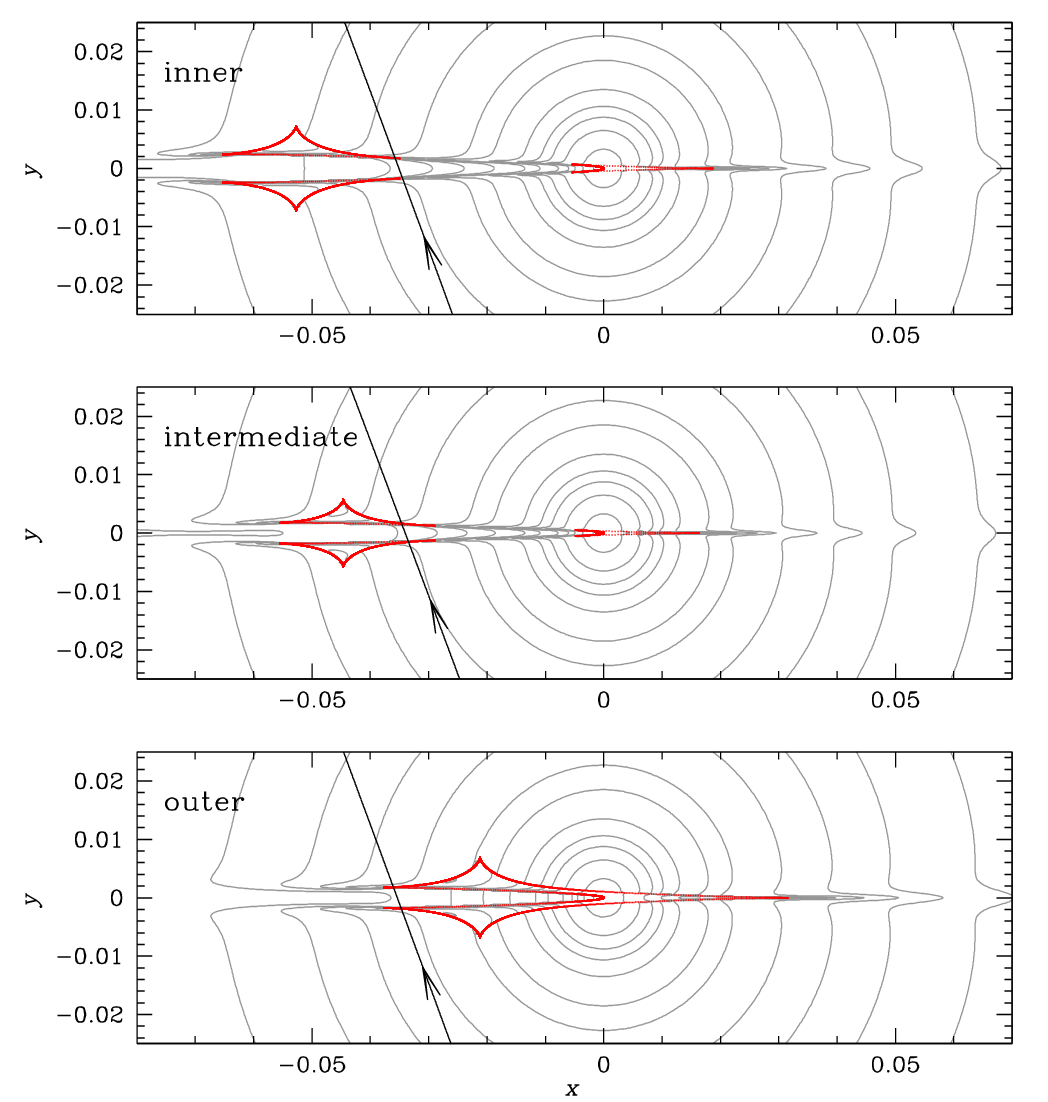}
\caption{
Lens system configurations of the three local solutions: "inner", "intermediate", and "outer"
solutions. In each panel, the figures drawn in red represent the caustic, and the line with 
an arrow denotes the source trajectory. The curves surrounding the caustics indicate the 
equi-magnification contours.  The coordinates are centered at the position of the planet host.
}
\label{fig:four}
\end{figure}

Figure~\ref{fig:two} displays the $\Delta\chi^2$ map on the $(\log s, \log q)$ parameter plane 
obtained from the grid search.  The lower map shows entire grid of the inspected region, while 
the upper map displays a closer look at the area surrounding local solutions.  From the 
investigation of the map, we identified three local solutions, which centered at $(\log s, 
\log q) \sim (-0.015, -4.55)$, $\sim (-0.009, -4.90)$, and $\sim (-0.001, -4.55)$. We assign 
the individual solutions the labels of "inner," "intermediate", and "outer," with the rationale 
behind these designations to be discussed below. Regardless of the solutions, the estimated mass 
ratios are $q < 10^{-4}$, suggesting that the anomaly was caused by a companion of very low mass 
associated with the lens.

Table~\ref{table:two} presents the complete lensing parameters for the three identified 
local solutions. The parameters of these solutions were refined from those obtained from the 
grid search by allowing unrestricted variation across all parameters and by considering 
higher-order effects causing deviations of the relative lens-source motion from rectilinear. 
We examined two distinct higher-order effects: the first arises from Earth's orbital motion 
around the Sun, known as microlens-parallax effects \citep{Gould1992a}, while the second stems 
from the orbital motion of a planet around its host, referred to as lens-orbital effects 
\citep{Batista2011, Skowron2011}. For the consideration of the microlens-parallax effect, we 
added two lensing parameters $(\pien, \piee)$, which respectively denote the north and east 
components of the microlens-parallax vector $\pievec$. The microlens-parallax vector is 
related to the distance to the lens $\dl$ and source $\ds$, the relative parallax of the lens 
and source, $\pi_{\rm rel} = {\rm au}(1/\dl - 1/\ds)$, and the relative lens-source proper 
motion by
\begin{equation}
\pievec = \left( {\pi_{\rm rel} \over \thetae} \right) \left( {\muvec \over \mu} \right).
\label{eq1}
\end{equation}
Under the assumption of a minor change in the lens configuration, we account for the lens-orbital
effect by introducing two additional parameters $(ds/dt, d\alpha/dt)$, which represent the yearly 
rates of change in the binary separation and the angle of the source trajectory, respectively. 
Upon comparing the fits, it was observed that the degeneracies among the solutions are very 
severe with $\Delta\chi^2 < 1.0$.

In Figure~\ref{fig:three}, we present the model curves of the three local solutions and their 
residuals in the region around the anomaly. The models are so alike that they are barely 
distinguishable within the line width. For all solutions, the anomaly is described by a dip 
feature surrounded by shallow hills appearing on both sides of the dip. According to the models, 
the two KMTC data points around ${\rm HJD}^\prime \sim 229.5$ align with the dip feature's valley, 
while the two MOA points at ${\rm HJD}^\prime =229.89$ and 229.97 correspond to the right-side 
hill.

The lens-system configurations corresponding to the individual solutions are depicted in 
Figure~\ref{fig:four}.  Across all solutions, the source consistently traversed the negative 
deviation region located behind the caustic positioned around the planet host. However, its 
specific alignment relative to the caustic differs among solutions. In the inner solution, 
the caustic consists of three segments, with one segment positioned near the planet's host 
(central caustic) and the other two segments (the planetary caustic) situated away from the 
host, on the opposite side of the planet. The source trajectory passed through the inner area 
between these central and planetary caustics.  Conversely, within the outer solution, the 
merging of the central and planetary caustics creates a unified resonant caustic. The source 
path crossed the outer region of this caustic.  We classify these solutions as "inner" and 
"outer", determined by the side of the caustic that the source passed through. In the 
intermediate solution, the caustic configuration resembles that of the inner solution, yet 
in this case, the source trajectory crossed the right side of the planetary caustic, unlike 
the trajectory of the inner solution.

\begin{figure}[t]
\includegraphics[width=\columnwidth]{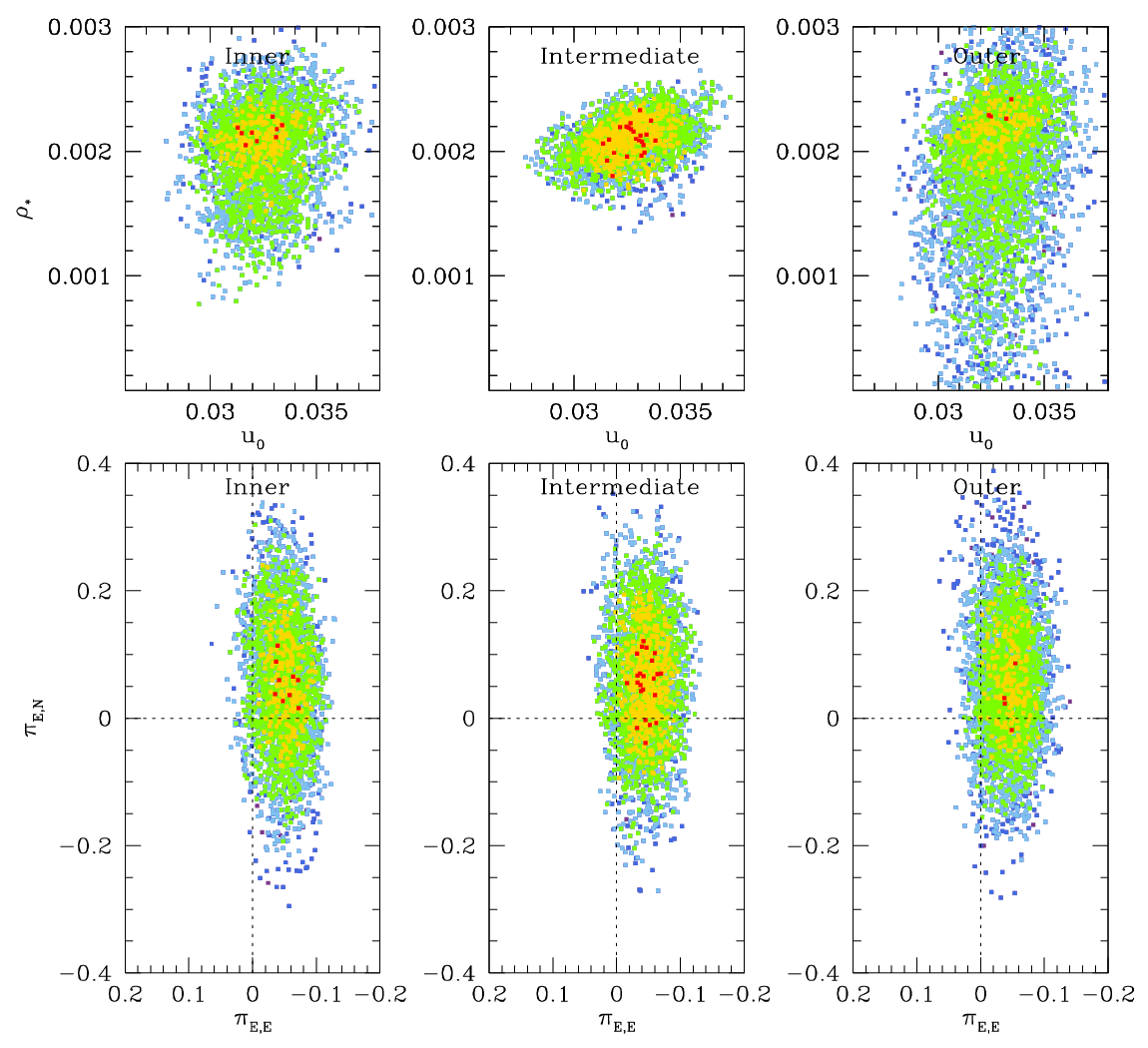}
\caption{
Scatter plots of points in the MCMC chain on the $u_0$--$\rho$ (upper panels) and $\piee$--$\pien$ 
(lower panels) parameter planes.  The color scheme matches that utilized in Fig.~\ref{fig:two}.
}
\label{fig:five}
\end{figure}

The normalized source radius is measured, but the associated uncertainty varies depending on the 
solutions, as illustrated in the three upper panels of Figure~\ref{fig:five}. These panels depict 
scatter plots of points in the MCMC chain on the $u_0$--$\rho$ parameter planes for the individual 
solutions. While the median values, $\langle \rho \rangle \sim 2 \times 10^{-3}$, exhibit 
consistency across solutions, there is notable diversity in the corresponding uncertainties. 
Specifically, the uncertainty ranges from $\sigma(\rho) \sim 0.18 \times 10^{-3}$ for the 
intermediate solution to $\sigma(\rho) \sim 0.61 \times 10^{-3}$ for the outer solution. This 
variation arises due to disparities in the lens system configurations among the solutions.

It turned out that the similarity between the model curves of the inner and outer solutions 
was caused by the inner--outer degeneracy. This degeneracy was initially proposed by 
\citet{Gaudi1997} to indicate the resemblance in planetary signals resulting from the 
source trajectories passing on near and far sides of a planetary caustic.  \citet{Yee2021}, 
\citet{Zhang2022a}, and \citet{Zhang2022b} later extended this concept to encompass planetary 
signals related to central and resonant caustics.  Subsequently, \citet{Hwang2022} and 
\citet{Gould2022} proposed an analytical relationship between the lensing parameters of the 
pair of solutions affected by this degeneracy:
\begin{equation}
s_\pm^\dagger = \sqrt{s_{\rm in} \times s_{\rm out}} = 
{\sqrt{u_{\rm anom}^2+4}\pm u_{\rm anom} \over 2}.
\label{eq2}
\end{equation}
Here 
$u^2_{\rm anom} = \tau^2_{\rm anom} + u_0^2$, $\tau_{\rm anom}=(t_{\rm anom}-t_0)/\te$, 
$t_{\rm anom}$ indicates the time of the anomaly, and $s_{\rm in}$ and $s_{\rm out}$ 
represent the planetary separations of the inner and outer solutions, respectively.  For 
anomalies showing a bump feature, the sign in the right term is $(+)$, whereas for those 
displaying a dip feature, the sign is $(-)$.  From the lensing parameters $(t_0, u_0, \te, 
s_{\rm in}, s_{\rm out}, t_{\rm anom})\sim (228.65, 3.3\times 10^{-2}, 75, 0.9740, 0.9894, 
229.57)$, we find that the geometric mean $\sqrt{s_{\rm in} \times s_{\rm out}} = 0.982$ 
matches very well the value $\left(\sqrt{u_{\rm anom}^2+4}\pm u_{\rm anom} \right)/2=0.983$.

Although the fit improvement from the static model is marginal, the microlens-parallax parameters 
could be constrained.  This is demonstrated in the scatter plot of MCMC points within the $(\piee, 
\pien)$ parameter plane, showcased in the lower panels of Figure~\ref{fig:five}.  Measuring these 
parameters is important because they offer constraints on the physical parameters of the lens 
\citep{Gould1992a, Gould2000}.  In contrast, the determined orbital parameters exhibit considerable 
uncertainties.

\begin{figure}[t]
\includegraphics[width=\columnwidth]{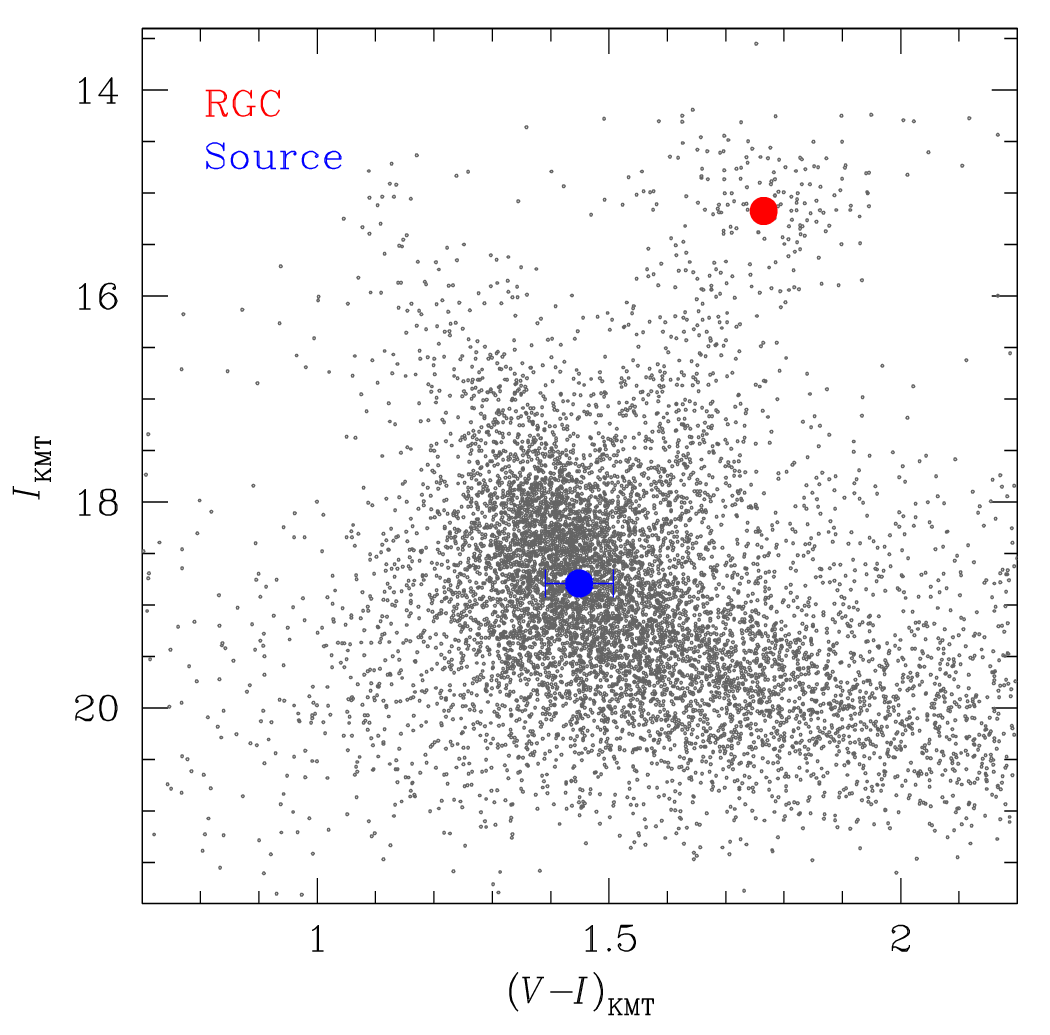}
\caption{
Positions of the source (blue dot) and red giant clump (RGC: red dot) centroid in the instrumental 
color-magnitude diagram of stars adjacent to the source star. 
}
\label{fig:six}
\end{figure}

\section{Source star and constraint on Einstein radius}\label{sec:four}

In this section, we define the source star of the event. Defining the source is important not only
to to fully characterize the event but also to estimate the angular Einstein radius. The value of
Einstein radius is constrained from the angular source radius by the relation
\begin{equation}
\thetae = {\theta_* \over \rho},
\label{eq3}
\end{equation}
where angular source radius $\theta_*$ is deduced from the stellar type of the source.

The characterization of the source follows the methodology outlined in \citet{Yoo2004}. Initially,
we constructed the $I$ and $V$-band light curves of the event using the pyDIA code \citep{Albrow2017}. 
Subsequently, we determined the flux values of the source, $F_{\rm S}$, in the individual passbands
by regressing the pyDIA light curves with respect to the lensing model, $A_{\rm model}(t)$, that is,
\begin{equation}
F_{\rm obs}(t) = A_{\rm model}(t) F_{\rm S}(t) + F_b.
\label{eq4}
\end{equation}
Here, $F_{\rm obs}(t)$ represents the observed flux of the event and $F_b$ represents the flux 
contributed by blended stars. In the next step, we positioned the source in the instrumental 
color-magnitude diagram (CMD), which is constructed from the pyDIA photometry of stars located 
near the source.  Finally, we converted the instrumental source color and magnitude, denoted as 
$(V-I, I)_{\rm S}$, into de-reddened values, $(V-I, I)_{{\rm S},0}$, utilizing the centroid of the 
red giant clump (RGC), with its instrumental color and magnitude $(V-I, I)_{\rm RGC}$, in the CMD 
as a reference, that is, 
\begin{equation}
(V-I, I)_{{\rm S},0} = (V-I, I)_{{\rm RGC},0} + \Delta(V-I, I). 
\label{eq5}
\end{equation}
Here, $\Delta(V-I, I) = (V-I, I)_{\rm S} - (V-I, I)_{\rm RGC}$ represents the offsets in color and 
magnitude of the source from those of the RGC centroid, and $(V-I, I)_{{\rm RGC},0} = (1.060, 14.339)$ 
denote the de-reddened values of the RGC centroid, as determined by \citet{Bensby2013} and \citet{Nataf2013}.

\begin{table}[t]
\caption{Source parameters.\label{table:three}}
\begin{tabular*}{\columnwidth}{@{\extracolsep{\fill}}lllcc}
\hline\hline
\multicolumn{1}{c}{Parameter}          &
\multicolumn{1}{c}{Value}              \\
\hline
 $(V-I, I)_{\rm S}      $     &  $(1.449 \pm 0.059, 18.791 \pm 0.042)$     \\ 
 $(V-I, I)_{\rm RGC}    $     &  $(1.765, 15.173)                    $     \\ 
 $(V-I, I)_{{\rm RGC},0}$     &  $(1.060, 14.339)                    $     \\ 
 $(V-I, I)_{{\rm S},0}  $     &  $(0.744 \pm 0.071, 17.957 \pm 0.046)$     \\ 
\hline
\end{tabular*}
\end{table}

In Figure~\ref{fig:six}, we mark the positions of the source (blue dot) and RGC centroid (red dot) 
in the instrumental CMD.  The CMD is constructed from the pyDIA photometry of stars in the KMTC image 
lying $3.4 \times 3.4$~arcmin$^2$ around the source.  The estimated values of $(V-I, I)_{\rm S}$, 
$(V-I, I)_{\rm RGC}$, and $(V-I, I)_{{\rm S},0}$ are detailed in Table~\ref{table:three}. Based on 
the de-reddened color and magnitude, the source is identified as a turnoff star of a mid G spectral 
type. For the estimation of the angular source radius, we first converted the measured $V-I$ color 
into $V-K$ color with the use of the \citet{Bessell1988} relation, and then deduced $\theta_*$ from 
the \citet{Kervella2004} relation between $(V-K, I)$ and $\theta_*$. The estimated angular source 
radius from this process is
\begin{equation}
\theta_* = 0.837 \pm 0.083~\mu{\rm as}. 
\label{eq6}
\end{equation}
When combined with the normalized source radii associated with the distinct degenerate solutions 
presented in Table~\ref{table:two}, this provides the Einstein radii corresponding to the 
individual solutions of
\begin{equation}
\theta_{\rm E} =  {\theta_* \over \rho} = 
\begin{cases}
	0.367 \pm 0.070~{\rm mas}    & \text{(inner)},        \\
	0.420 \pm 0.057~{\rm mas}    & \text{(intermediate)}, \\
	0.371 \pm 0.107~{\rm mas}    & \text{(outer)},        \\
\end{cases}
\label{eq7}
\end{equation}
and the values of relative lens-source proper motion of
\begin{equation}
\mu =  {\theta_{\rm E} \over \te} = 
\begin{cases}
1.79 \pm 0.34~{\rm mas/yr}    & \text{(inner)},        \\
1.99 \pm 0.27~{\rm mas/yr}    & \text{(intermediate)}, \\
1.82 \pm 0.52~{\rm mas/yr}    & \text{(outer)}.        \\
\end{cases}
\label{eq8}
\end{equation}

As we will describe, the blended light (formally $f_{\rm b}=0.038\pm 0.019$) is not reliably detected.
That is, it is consistent with zero. We can use this non-detection to place limits on lens light.
There are three sources of uncertainty in the measurement of the blend light. The first is the
formal uncertainty from the fit, in which the baseline flux is formally treated as perfectly
measured: $\sigma_{\rm formal} = 0.019$. The second is the photon error in the measurement
of the baseline flux $\sigma_{\rm base} = 0.025$. The quadrature sum of these is $\sigma_{\rm
naive} = (\sigma_{\rm formal}^2 + \sigma_{\rm base}^2)^{1/2} = 0.031$, which is equivalent to
an $I_{\rm naive} = 21.8$ star. This is far below the error due to the mottled background of
unresolved stars \citep{Park2004}. Because this field is far from the Galactic plane and center 
and so is very sparse, we conservatively estimate the $3\,\sigma$ upper limit on the lens flux 
due to mottled background as $f_L < 0.25$ or
\begin{equation}
I_{\rm L} > 19.5.
\label{eq9}
\end{equation}
We note that this constraint eliminates roughly half of lens distribution that would be obtained in
its absence, and thus it is highly significant.

\begin{figure}[t]
\includegraphics[width=\columnwidth]{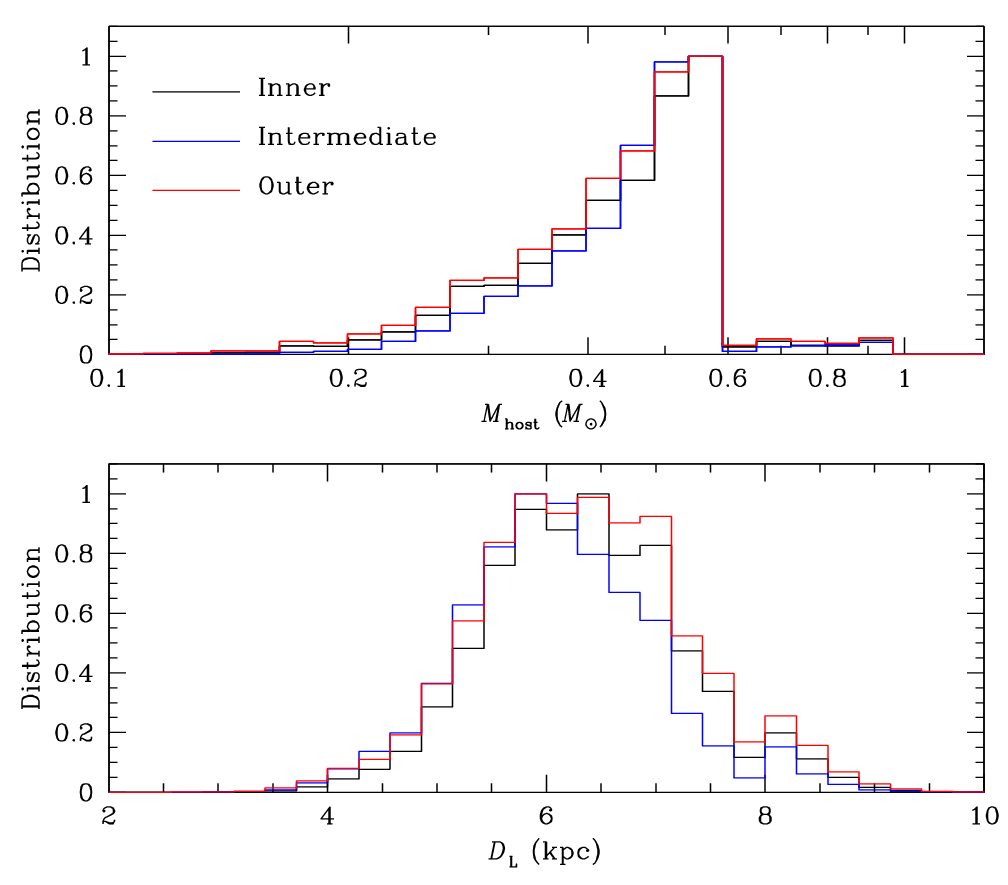}
\caption{
Bayesian posteriors of the primary lens mass ($M_{\rm host}$) and distance to the lens ($\dl$).
In each panel, three curves drawn in black, blue, and red represent posteriors corresponding to 
the inner, intermediate, and outer solutions, respectively. 
}
\label{fig:seven}
\end{figure}

\section{Physical lens parameters}\label{sec:five}

We determine the physical lens parameters through a Bayesian analysis that integrates constraints 
from measured lensing observables with priors derived from the physical and dynamical distributions, 
along with the mass function of lens objects within the Milky Way. This analysis begins by generating 
numerous synthetic events via Monte Carlo simulation. Each event's mass ($M_i$) was derived from a 
model mass function, and the distances  to the lens and source ($D_{{\rm L},i}$, $D_{{\rm S},i}$), 
alongside their relative proper motion ($\mu_i$), were inferred using a Galaxy model. Our simulation 
incorporates the mass function model proposed by \citet{Jung2018} and utilizes the Galaxy model 
introduced in \citet{Jung2021}.  The bulge density profile in the Galaxy model conforms to the triaxial 
model described by \citet{Han1995}, while the density profile of disk objects adheres to the modified 
double-exponential form as presented in Table 3 of their paper.  In the mass function, we did not include 
stellar remnants because the formation of planets around stellar remnants is considered less probable 
due to disruptive events associated with these stellar objects. Subsequently, we calculated the time 
scale and Einstein radius of each synthetic event using relations represented by
\begin{equation}
t_{{\rm E},i} = {\theta_{{\rm E},i} \over \mu_i}; \qquad
\theta_{{\rm E},i} = \sqrt{\kappa M_i \pi_{{\rm rel},i}},
\label{eq10}
\end{equation}
where $\kappa = 4G/(c^2{\rm au}) = 8.14~{\rm mas}/M_\odot$.  The microlens parallax was computed 
using the relation presented in Eq.~(\ref{eq1}). In the subsequent step, we assigned a weight to 
each event proportional to
\begin{equation}
w_i = \exp \left(  {\chi_i^2 \over 2}\right),
\label{eq11}
\end{equation}
where $\chi_i^2$ value was computed using the relation
\begin{equation}
\eqalign{
\chi_i^2  = &
{ (t_{{\rm E},i}-\te ) \over \sigma^2(\te)} +
{ (\theta_{{\rm E},i}-\thetae ) \over \sigma^2(\thetae)} \cr
 & + 
 \sum_{j=1}^2  \sum_{k=1}^2
b_{j,k}
(\pi_{{\rm E},j,i} - \pi_{{\rm E},i})
(\pi_{{\rm E},k,i} - \pi_{{\rm E},i}).
}
\label{eq12}
\end{equation}
Here, $[\te, \sigma(\te)]$ and  $[\thetae, \sigma(\thetae)]$ denote the measured time scale and 
Einstein radius and their uncertainty, respectively, and $b_{j,k}$ represents the inverse covariance 
matrix of the microlens-parallax vector $\pievec$, $(\pi_{{\rm E},1}, \pi_{{\rm E},2})_i = (\pien, 
\piee)_i$ represent the parallax parameters of each simulated event, while $(\pien, \piee)$ denotes 
the measured parallax parameters.

\begin{table}[t]
\caption{Physical lens parameters.\label{table:four}}
\begin{tabular*}{\columnwidth}{@{\extracolsep{\fill}}llll}
\hline\hline
\multicolumn{1}{c}{Parameter}          &
\multicolumn{1}{c}{Inner}              &
\multicolumn{1}{c}{Intermediate}       &
\multicolumn{1}{c}{Outer}              \\  
\hline
  $M_{\rm host}$ ($M_\odot$)      &  $0.49 ^{+0.10}_{-0.15}$     &   $0.51 ^{+0.08}_{-0.13}$     &  $0.48^{+0.10}_{-0.15}$     \\  [0.8ex]
  $M_{\rm planet}$ ($M_{\rm E}$)  &  $3.98 ^{+0.78}_{-1.25}$     &   $2.63 ^{+0.42}_{-0.68}$     &  $3.69^{+0.77}_{-1.18}$     \\  [0.8ex]
  $\dl$ (kpc)                     &  $6.47 ^{+0.88}_{-0.85}$     &   $6.21 ^{+0.87}_{-0.80}$     &  $6.47 ^{+0.95}_{-0.91}$    \\  [0.8ex]
  $a_\perp$ (au)                  &  $2.48 ^{+0.34}_{-0.32}$     &   $2.63 ^{+0.37}_{-0.34}$     &  $2.50 ^{+0.37}_{-0.35}$    \\  [0.8ex]
\hline                                                                                    
\end{tabular*}                                                                            
\end{table}                                                                              

Aside from the constraint from the lensing observables, we incorporate an additional
constraint derived from the blended flux. This constraint is rooted in the relationship 
between the lens flux and the overall blended flux, necessitating the lens flux to be lower 
than the total blending flux. To compute the $I$-band magnitude of the lens, we utilize the 
equation:
\begin{equation}
I_{\rm L} = M_{I,{\rm L}} + 5 \log \left( {\dl\over {\rm pc}} \right) - 5 + A_{I,{\rm L}}. 
\label{eq13}
\end{equation}
Here, $M_{I,{\rm L}}$ denotes the absolute magnitude of the lens, and $A_{I,{\rm L}}$ 
represents the extinction at a distance $\dl$.  For the derivation of absolute magnitude 
$M_{I,{\rm L}}$ corresponding the lens mass, we used the \citet{Pecaut2013} mass--luminosity 
relation, which was derived from pre-main sequence stars in nearby, negligibly reddened stellar 
groups.  The extinction is modeled as
\begin{equation}
A_{I,{\rm L}} = A_{I,{\rm tot}} 
\left[ 1-\exp -\left( {|z|\over h_{z,{\rm dust}}}\right) \right],
\label{eq14}
\end{equation}
where $A_{I,{\rm tot}} = 0.46$ indicates the overall extinction observed in the field, 
$h_{z,{\rm dust}} = 100$~pc represents the dust's scale height, $z = \dl \sin b + z_0$
and $z_0 = 15$~pc denote the heights of the lens and the Sun above the Galactic plane, 
respectively.  We note that dust may be distributed unevenly in patches, and therefore our 
model, which assumes a smooth distribution, provides only an approximate description of 
the dust.  We enforced the blending constraint by assigning a weight  $w_i=0$ to artificial 
events featuring lenses whose brightness fails to meet the condition specified in Eq.~(\ref{eq9}).  
It turns out that the blending constraint has a significant impact on the estimation of lens 
parameters.

Figure~\ref{fig:seven} shows the posteriors of the primary lens mass and distance to the 
lens.  Table~\ref{table:four} presents the estimated host mass ($M_{\rm host}$), planet 
mass ($M_{\rm planet}$), distance ($\dl$) to the planetary system, and projected planet-host 
separation ($a_\perp=s\thetae \dl$) corresponding to the individual degenerate solutions.  
The median of each posterior distribution was chosen as the representative value, while the 
uncertainties were quantified by the 16\% and 84\% of the distribution.  Based on the estimated 
mass, the host of the planet is identified as a low-mass star of an early M spectral type.
The estimated mass of the planet falls within the range of approximately 2.6 to 4.0 times Earth's 
mass in terms of the median value, signifying a notably low mass.  Positioned at a distance of 
about (6.2--6.5)~kpc from Earth, this planetary system resides in a region where approximately 
85\% of the stellar population belongs to the Galactic bulge, with the remaining 15\% in the disk.  
The projected separation between the planet and its host star is approximately 2.5~au, roughly 
two times farther than the ice line distance.

\begin{figure}[t]
\includegraphics[width=\columnwidth]{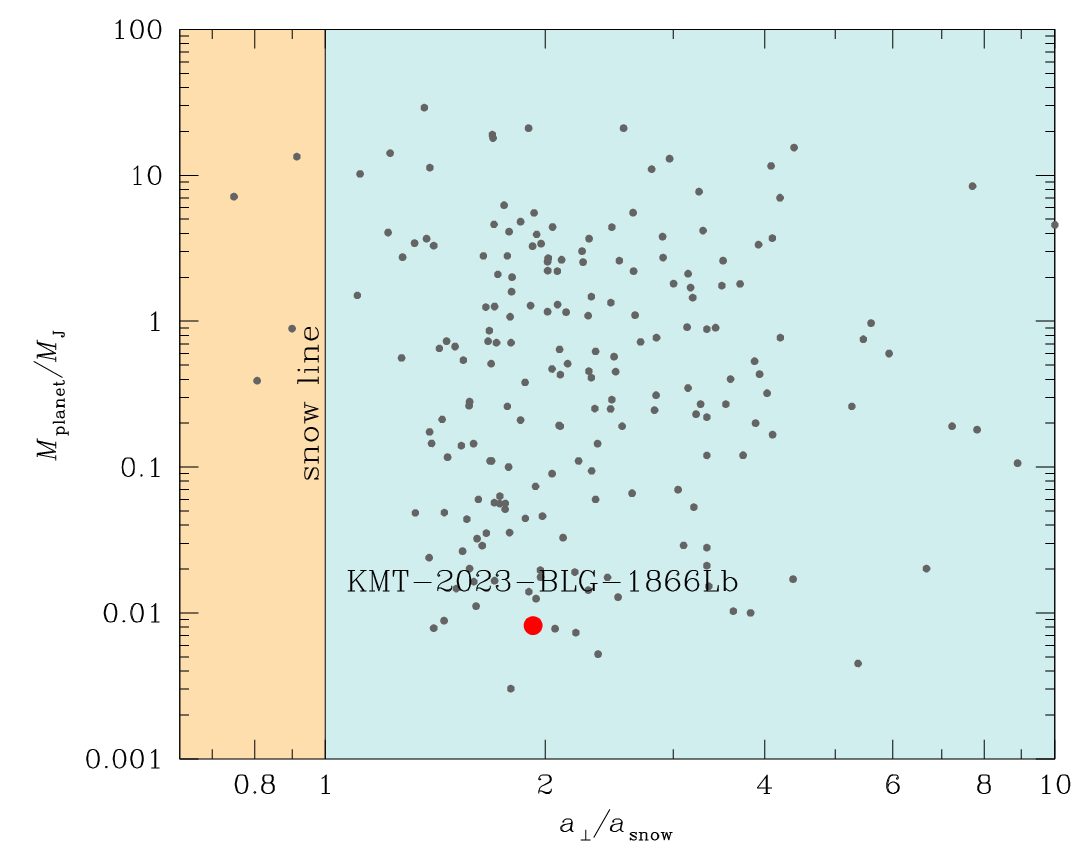}
\caption{
Distribution of microlensing planets in the parameter space defined by their mass and 
projected separation.  The separation is scaled to the snow line of planet host, which is 
represented by a vertical line. The red dot indicates the location of KMT-2023-BLG-1866Lb.
}
\label{fig:eight}
\end{figure}

The identification of planet KMT-2023-BLG-1866Lb is of significant scientific importance 
due to its remarkably low mass. Despite the growing planet detection efficiency of lensing 
experiments, planets falling into the categories of terrestrial planets (with masses 
$\lesssim 2~M_{\rm E}$) and super-Earths (with masses around $\sim 3$ -- 10 $M_{\rm E}$) 
still represent a small fraction of the microlensing planets. KMT-2023-BLG-1866Lb occupies 
a region of very low mass where planets are sparsely distributed, as depicted in 
Figure~\ref{fig:eight}.  Therefore, its detection signifies the extension of the lensing 
surveys to very low-mass planets that have been difficult to be detected from the earlier 
surveys.

\section{Summary}\label{sec:six}

We investigated the nature of the brief anomaly observed close to the peak of the lensing light
curve in KMT-2023-BLG-1866. This anomaly was partially covered, primarily because of its short
duration together with cloudy weather conditions at the observation sites and limited nighttime
availability toward the end of the bulge season.  To address the inherent challenges of interpreting 
partially covered signals, we conducted a rigorous exploration of all plausible degenerate solutions. 
This in-depth analysis revealed three distinct planetary scenarios that provide equally valid 
explanations for the observed anomaly. These scenarios are characterized by the specific planetary 
parameters: 
$(s, q)_{\rm inner} = [0.9740 \pm 0.0083, (2.46 \pm 1.07) \times 10^{-5}]$, 
$(s, q)_{\rm intermediate} = [0.9779 \pm 0.0017, (1.56 \pm 0.25)\times 10^{-5}]$, and 
$(s, q)_{\rm outer} = [0.9894 \pm 0.0107, (2.31 \pm 1.29)\times 10^{-5}]$.
We identified that the ambiguity between the
inner and outer solutions stems from the inner-outer degeneracy, while the similarity between 
the intermediate solution and the others is due to an accidental degeneracy caused by incomplete
anomaly coverage.  Through a Bayesian analysis utilizing the constraints derived from measured 
lensing observables and blending flux, our estimation indicates that the lens system comprises 
a very low-mass planet orbiting an early M-type star situated approximately (6.2 -- 6.5) kpc from 
Earth in terms of median posterior values for the different solutions.  The median mass of the 
planet host is in the range of (0.48 -- 0.51)~$M_\odot$, and that of the planet's mass spans a 
range of (2.6 -- 4.0)~$M_{\rm E}$, varying across different solutions.

\begin{acknowledgements}
Work by C.H. was supported by the grants of National Research Foundation of Korea (2019R1A2C2085965). 
J.C.Y. and I.-G.S. acknowledge support from U.S. NSF Grant No. AST-2108414. 
Y.S. acknowledges support from BSF Grant No. 2020740.
This research has made use of the KMTNet system operated by the Korea Astronomy and Space
Science Institute (KASI) at three host sites of CTIO in Chile, SAAO in South Africa, and SSO in
Australia. Data transfer from the host site to KASI was supported by the Korea Research
Environment Open NETwork (KREONET). This research was supported by KASI under the R\&D
program (Project No. 2024-1-832-01) supervised by the Ministry of Science and ICT.
W.Z. and H.Y. acknowledge support by the National Natural Science Foundation of China (Grant
No. 12133005).
W. Zang acknowledges the support from the Harvard-Smithsonian Center for Astrophysics through
the CfA Fellowship. 
The MOA project is supported by JSPS KAKENHI Grant Number JP24253004, JP26247023,JP16H06287 and JP22H00153.
\end{acknowledgements}


\begin{thebibliography}{}
\bibitem[Albrow et al.(2009)]{Albrow2009} Albrow, M., Horne, K., Bramich, D.~M., et al.\ 2009, \mnras, 397, 2099
\bibitem[Albrow(2017)]{Albrow2017} Albrow, M.\ 2017, MichaelDAlbrow/pyDIA: Initial Release on Github,Versionv1.0.0, Zenodo, doi:10.5281/zenodo.268049
\bibitem[Alcock et al.(1993)]{Alcock1993} Alcock, C., Akerlof, C. W., Allsman, R. A., et al. 1993, Nature, 365, 621
\bibitem[An \& Han(2002)]{An2002} An, J. H., \& Han, C. 2002, \apj, 573, 351
\bibitem[Aubourg et al.(1993)]{Aubourg1993} Aubourg, E., Bareyre, P., Br\'ehin, S., et al. 1993, Nature, 365, 623
\bibitem[Batista et al.(2011)]{Batista2011} Batista, V., Gould, A., Dieters, S. et al. 2011, \aap, 529, 102
\bibitem[Beaulieu et al.(2006)]{Beaulieu2006} Beaulieu, J.-P., Bennett, D. P., Fouqu\'e, P., et al. 2006, Nature, 439, 437
\bibitem[Bell et al.(2023)]{Bell2023} Bell, A., Zhang, J., Jung, Y. K., et al. 2023, \mnras, submitted (arXiv:2311.13097)
\bibitem[Bensby et al.(2013)]{Bensby2013} Bensby, T., Yee, J.~C., Feltzing, S., et al.\ 2013, \aap, 549, A147
\bibitem[Bessell \& Brett(1988)]{Bessell1988} Bessell, M. S., \& Brett, J. M. 1988, \pasp, 100, 1134
\bibitem[Bond et al.(2001)]{Bond2001} Bond, I. A., Abe, F., Dodd, R. J., et al. 2001, \mnras, 327, 868
\bibitem[Bond et al.(2004)]{Bond2004} Bond, I. A., Udalski, A., Jaroszy\'nski, M., et al. 2004, \apj, 606, L155
\bibitem[Choi et al.(2012)]{Choi2012} Choi, J.-Y., Shin, I.-G., Han, C., et al. 2012, \apj, 756, 48 
\bibitem[Claret(2000)]{Claret2000} Claret, A. 2000, \aap, 363, 1081
\bibitem[Dong et al.(2006)]{Dong2006} Dong, S., DePoy, D. L., Gaudi, B. S., et al. 2006, \apj, 642, 842
\bibitem[Doran \& Mueller(2004)]{Doran2004} Doran, M., \& Mueller, C. M. 2004, J. Cosmology Astropart. Phys., 09, 003
\bibitem[Gaudi  \& Gould(1997)]{Gaudi1997} Gaudi, B. S., \& Gould, A. 1997, \apj, 486, 85
\bibitem[Gaudi(1998)]{Gaudi1998} Gaudi, B. S. 1998, \apj, 506, 533
\bibitem[Gould(1992)]{Gould1992a} Gould, A. 1992, \apj, 392, 442
\bibitem[Gould \& Loeb(1992)]{Gould1992b}  Gould, A., \& Loeb, A. 1992, \apj, 396, 104
\bibitem[Gould(2000)]{Gould2000} Gould, A. 2000, \apj, 542, 785
\bibitem[Gould et al.(2006)]{Gould2006} Gould, A., Udalski, A., An, D., et al. 2006, \apjl, 644, L37
\bibitem[Gould et al.(2014)]{Gould2014} Gould, A., Udalski, A., Shin, I.-G., et al. 2014, Science, 345, 46
\bibitem[Gould et al.(2020)]{Gould2020} Gould, A., Ryu, Y.-H., Calchi Novati, S., et al. 2020, JKAS, 53, 9
\bibitem[Gould et al.(2022)]{Gould2022} Gould, A., Han, C., Zang, W., et al. 2022, \aap, 664, A13
\bibitem[Griest \& Safizadeh(1998)]{Griest1998} Griest, K., \& Safizadeh, N. 1998, \apj, 500, 37
\bibitem[Han \& Gould(1995)]{Han1995} Han, C., \& Gould, A. 1995, \apj, 447, 53
\bibitem[Han et al.(2021)]{Han2021} Han, C., Udalski, A., Lee, C.-U., et al. 2021, \aap, 649, A90
\bibitem[Han et al.(2022a)]{Han2022a} Han, C., Gould, A., Albrow, M. D., et al. 2022a, \aap, 658, A62
\bibitem[Han et al.(2022b)]{Han2022b} Han, C., Bond, I. A., Yee, J. C., et al. 2022b, \aap, 658, A94
\bibitem[Han et al.(2022c)]{Han2022c} Han, C., Kim, D., Gould, A., et al. 2022c, \aap, 664, A33
\bibitem[Han et al.(2023a)]{Han2023a} Han, C., Gould, A., Jung, Y. K., et al. 2023a, \aap, 674, A89
\bibitem[Han et al.(2023b)]{Han2023b} Han, C., Lee, C.-U., Zang, W., et al. 2023b, \aap, 674, A90
\bibitem[Herrera-Mart\'in et al.(2020)]{Herrera2020} Herrera-Mart\'in, A., Albrow, M. D., Udalski, A. et al. 2020, \aj, 159, 256
\bibitem[Hwang et al.(2018a)]{Hwang2018a} Hwang, K. H., Kim, H. W., Kim, D. J., et al. 2018a, JKAS, 51, 197
\bibitem[Hwang et al.(2018b)]{Hwang2018b} Hwang, K.-H., Udalski, A., Shvartzvald, Y., et al. 2018b, \aj, 155, 20
\bibitem[Hwang et al.(2022)]{Hwang2022} Hwang, K.-H., Zang, W., Gould, A., et al. 2022, \aj, 163, 43
\bibitem[Jung et al.(2018)]{Jung2018} Jung, Y. K., Udalski, A., Gould, A., et al. 2018, \aj, 155, 219
\bibitem[Jung et al.(2020)]{Jung2020} Jung, Y. K., Udalski, A., Zang, W., et al. 2020, \aj, 160, 255
\bibitem[Jung et al.(2021)]{Jung2021} Jung, Y. K., Han, C., Udalski, A., et al. 2021, \aj, 161, 293
\bibitem[Kervella et al.(2004)]{Kervella2004} Kervella, P., Th\'evenin, F., Di Folco, E., \& S\'egransan, D.\ 2004, \aap, 426, 29
\bibitem[Kim et al.(2016)]{Kim2016} Kim, S.-L., Lee, C.-U., Park, B.-G., et al.\ 2016, JKAS, 49, 37
\bibitem[Kondo et al.(2021)]{Kondo2021} Kondo, I., Yee, J. C., Bennett, D. P., et al. 2021, \aj, 162, 77
\bibitem[Mao \& Paczy\'nski(1991)]{Mao1991} Mao, S. \& Paczy\'nski, B. 1991, \apj, 374, L37
\bibitem[Muraki et al.(2011)]{Muraki2011} Muraki, Y., Han, C., Bennett, D. P., et al. 2011, \apj, 741, 22
\bibitem[Nataf et al.(2013)]{Nataf2013} Nataf, D.~M., Gould, A., Fouqu\'e, P., et al.\ 2013, \apj, 769, 88
\bibitem[Paczy\'nski(1986)]{Paczynski986} Paczy\'nski, B. 1986, \apj, 304, 1
\bibitem[Park et al.(2004)]{Park2004} Park, B. -G., DePoy, D. L., Gaudi, B. S., et al. 2004, \apj, 609, 166 
\bibitem[Pecaut \& Mamajek(2013)]{Pecaut2013} Pecaut, M. J., \& Mamajek, E. E. 2013, \apjs, 208, 9
\bibitem[Ranc et al.(2019)]{Ranc2019} Ranc, C., Bennett, D. P., Hirao, Y., et al. 2019, \aj, 157, 232
\bibitem[Robin et al.(2003)]{Robin2003} Robin, A. C., Reyl e, C., Derri\'ere, S., \& Picaud, S. 2003, \aap, 409, 523
\bibitem[Ryu et al.(2020)]{Ryu2020} Ryu, Y.-H., Udalski, A., Yee, J. C., et al. 2020, \aj, 160, 183
\bibitem[Ryu et al.(2022)]{Ryu2022} Ryu, Y.-H., Jung, Y. K., Yang, H., et al. 2022, \aj, 164, 180
\bibitem[Sumi et al.(2010)]{Sumi2010} Sumi, T., Bennett, D. P., Bond, I. A., et al. 2010, \apj, 710, 1641
\bibitem[Shvartzvald et al.(2017)]{Shvartzvald2017} Shvartzvald, Y., Yee, J. C., Novati, S. C., et al. 2017, \apjl, 840, L3
\bibitem[Skowron et al.(2011)]{Skowron2011} Skowron, J., Udalski, A., Gould, A., et al. 2011, \apj, 738, 87
\bibitem[Udalski et al.(1994)]{Udalski1994} Udalski, A., Szyma\'nski, M., Ka{\l}u\.zny, J., et al. 1994, Acta Astron., 44, 1
\bibitem[Udalski(2003)]{Udalski2003} Udalski, A. 2003, Acta Astron., 53, 291
\bibitem[Udalski et al.(2015)]{Udalski2015} Udalski, A., Szyma\'nski, M. K., Szyma\'nski, G., et al. 2015, Acta Astron., 65, 1
\bibitem[Udalski et al.(20118)]{Udalski2018} Udalski, A., Ryu, Y.-H., Sajadian, S., et al. 2018, Acta Astron., 68, 1
\bibitem[Yang et al.(2022)]{Yang2022} Yang, H., Zang, W., Gould, A., et al. 2022, \mnras, 516, 1894
\bibitem[Yang et al.(2024)]{Yang2024} Yang, H., Yee, J. C., Hwang, K.-H., et al. 2024, \mnras, 528, 11
\bibitem[Yee et al.(2012)]{Yee2012} Yee, J. C., Shvartzvald, Y., Gal-Yam, A., et al. 2012, \apj, 755, 102
\bibitem[Yee et al.(2021)]{Yee2021} Yee, J. C., Zang, W., Udalski, A., et al. 2021, \aj, 162, 180
\bibitem[Yoo et al.(2004)]{Yoo2004} Yoo, J., DePoy, D. L., Gal-Yam, A. et al. 2004, \apj, 603, 139
\bibitem[Zang et al.(2021a)]{Zang2021a} Zang, W., Hwang, K.-H., Udalski, A., et al. 2021a, \aj, 162, 163
\bibitem[Zang et al.(2021b)]{Zang2021b} Zang, W., Han, C., Kondo, I., et al. 2021b, Res. Astro. and Astroph., 21, 239
\bibitem[Zang et al.(2023)]{Zang2023} Zang, W., Jung, Y. K., Yang, H., et al. 2023, \aj, 165, 103
\bibitem[Zhang et al.(2022a)]{Zhang2022a} Zhang, K., Gaudi, B. S., \& Bloom, J. S. 2022, Nat Astron, 6., 782 
\bibitem[Zhang \& Gaudi(2022b)]{Zhang2022b} Zhang, K., \& Gaudi, B. S. 2022, \apjl, 936, L22 
\bibitem[Zhang et al.(2023)]{Zhang2023} Zhang, J., Zang, W., Jung, Y. K., et al. 2023, \mnras, 522, 6055
\end{thebibliography}
\end{document}